\documentclass[11pt]{article}
\usepackage[english]{babel}
\usepackage[utf8]{inputenc}
\usepackage{fancyhdr}
\usepackage{graphicx}
\usepackage{float}
\usepackage{array}
\usepackage{url}
\usepackage{xcolor}
\usepackage[paperheight=29.7cm,paperwidth=20.99cm,left=3.0cm,right=3.0cm,top=2.5cm,bottom=2.5cm]{geometry}

\title{Is the Solar System a Wilderness or a Construction Site? 
Conservationist and Constructivist Paradigms in Planetary Protection}

\author{Luk\'a\v{s} Likav\v{c}an$^{a}$$^{,}$$^{b}$$^{*}$ \\\\
        \small $^{a}$Researcher, Institute of Philosophy, Slovak Academy of Sciences \\
        \small $^{b}$Researcher, Antikythera, Berggruen Institute \\\\
        \small $^{*}$\tt{Email: lukas.likavcan@savba.sk, ORCID: 0000-0002-6957-9079}\\\\
}
\date{27 August 2025}

\begin{document}

\maketitle

\pagestyle{fancy}
\fancyhead[L]{\textit{Submitted to Acta Astronautica}}
\fancyhead[R]{\textit{Luk\'a\v{s} Likav\v{c}an}}
\fancyfoot[L]{\textbf{Article preprint version}}

\noindent{\textbf{Abstract: }Outer space exploration is one of the most prominent domains of earth-space governance. In this context, multiple policy documents by the UN, NASA, or Committee on Space Research (COSPAR) pledge to protect extraterrestrial environments from harmful human influence under the framework of \textit{planetary protection} or \textit{planetary stewardship}, understood primarily as the isolation of other celestial bodies from possible biological contamination. This paper analyses justifications of this framework that rely on analogies with the protection of terrestrial wilderness and nature’s intrinsic value, portraying them as representative of a \textit{conservationist paradigm} of earth-space governance. After presenting this paradigm, the paper builds an alternative \textit{constructivist paradigm}, grounded in recent findings about the evolution of the solar system, planets, and life. Ultimately, the paper argues that conservation is not the opposite of construction but one of its modalities: a conclusion that encourages the development of pragmatic protocols for space exploration instead of absolute imperatives.\\


\section{Introduction}

The intensification of human outer space activities in the ongoing
Second Space Age presents new governance challenges, since it is no
longer sufficient to think of outer space as a distant frontier, with
humanity as an occasional visitor {[}1{]}. This provokes the new
ramification of \emph{earth-space governance}, which ``challenges the
notion of the Earth system as a closed, isolated system and reckons the
expansion of human activities and ambitions into space'' {[}2{]}. Such statements are mainly motivated by the rapid acceleration of orbital
activities around the Earth, the increasing number of public and private
actors with launch capacities or space exploration agendas, and the
growing interest in extracting resources from other celestial bodies
(Moon, asteroids). The expansion in all these domains ``causes a flow of
influence that is no longer solely unidirectional from Earth to space
but also increasingly from space back to Earth,'' {[}2{]} thus imagining
a closed loop between Earth's environments and outer space, which
ultimately challenges the very boundary between the planet and its
cosmic environment.

This new ramification of earth-space governance is consistent with
contemporary philosophical elaborations that motivate a shift in human
attitude towards their planet. The recently emerged genre of
\emph{planetary thinking} (represented by recent works of Dipesh
Chakrabarty, Nigel Clark, or Bronislaw Szerszynski) treats the planet as
a dynamic cosmic entity rather than a stable ground of human existence
{[}3,4{]}. It can be best described as an open system, a radical
\emph{involution of cosmic exteriority} temporarily assembled into a
roughly spherical shape {[}5{]}. However, the established policy
frameworks and their implicit ethical paradigms struggle to keep pace
with this new perspective. The dominant theoretical approach grounding
the governance of human interaction with other worlds---a paradigm this
paper calls \emph{conservationist}---is rooted in an ontology of space
as a pristine, untouchable wilderness {[}6,7{]}. This approach underlies
concepts of \emph{planetary protection} and \emph{planetary
stewardship}, and the recent debates on their ethical or sustainability
dimensions {[}8{]}. As this paper argues, this approach---while
well-intentioned---may soon become philosophically untenable and
pragmatically unworkable. For this reason, the paper proposes a shift
from a conservationist to a \emph{constructivist} paradigm for
earth-space governance, particularly in the case of the solar system's
scientific exploration. By reframing the solar system not as a pristine
wilderness to be preserved but as a busy ``construction site'' of which
Earth and its life are the integral parts, the constructivist paradigm
offers a more philosophically robust and pragmatically flexible
framework for navigating the ethical complexities of space exploration.
This alternative paradigm does not simply grant license for
exploitation; rather, by recognizing humanity as an active agent within
cosmic evolution, it imposes a greater burden of responsibility to act
with foresight and measure its actions.

The paper is organized into five sections. Section 2 details the
existing policy proposals for the planetary protection regime,
emphasizing the issue of biological contamination present in policy
frameworks by key players in outer space exploration, such as NASA, the
UN Space Treaty, and the Committee on Space Research (COSPAR). Section 3
deconstructs the philosophical foundations of the conservationist
paradigm by exploring the metaphor of wilderness and the concept of
intrinsic value often invoked on the conservationist side of the debate.
Section 4 builds the alternative constructivist paradigm, grounding it
in recent scientific discoveries from sample return missions that reveal
the cosmic origins of life\textquotesingle s building blocks, as well as
the conceptual shifts in astrobiology towards more generous, inclusive
definitions of life. Section 5 translates this paradigm into potential
directives for action, arguing that conservation is not the opposite of
construction but one of its modalities. Finally, Section 6 concludes by
synthesising these arguments and exploring their broader implications
for outer space exploration.

\section{Planetary Protection}

As one of the four fundamental principles for the astrobiology research program, NASA’s 2008 Astrobiology Roadmap recognizes the value of \emph{planetary stewardship} “through an emphasis on protection against forward and back biological contamination and recognition of ethical issues associated with exploration” [9]. It thus subscribes to the spirit of the decades long history of \emph{planetary protection} efforts, starting with the founding of The Committee on Space Research (COSPAR) in 1958 and the subsequent establishing of the Planetary Quarantine Office at NASA in 1963, which underscores that the issue of cross-contamination between celestial bodies has been of an importance to the solar system’s exploration since its early days [10]. The 1967 UN Outer Space Treaty similarly emphasizes that 

\vspace{1\baselineskip}
\noindent"States Parties to the Treaty shall pursue studies of outer space, including the moon and other celestial bodies, and conduct exploration of them so as to avoid their harmful contamination and also adverse changes in the environment of the Earth resulting from the introduction of extraterrestrial matter and, where necessary, shall adopt appropriate measures for this purpose." [11]

\vspace{1\baselineskip}
\noindent As the major internationally recognized scientific organization promoting cooperation in outer space research, COSPAR has established its own Panel on Planetary Protection, with the Policy on Planetary Protection as its main document. This policy—adopted as the baseline reference by leading outer space science organizations such as NASA, ESA, and JAXA—distinguishes five categories of space missions according to their mission type (flyby, orbiter, probe, lander, Earth return) and target body (e.g. asteroids, Kuiper Belt Objects, gas giants, terrestrial planets, moons), ranked based on the probability of biological contamination (See \textbf{Table 1}). While the first four categories (I-IV) concern forward contamination (i.e. contamination of other celestial bodies by biological material from the Earth), the fifth category (V) accounts for backward contamination (i.e. contamination of earthly environments by extraterrestrial biological material).

\vspace{1\baselineskip}
\begin{center}
\renewcommand{\arraystretch}{1.5}
\noindent 
\begin{tabular}{|p{2cm}|p{4cm}|p{4cm}|p{3cm}|}
\hline
\textbf{Category} & \textbf{Type} & \textbf{Target body examples} & \textbf{Contamination risk} \\
\hline
I & All outbound types: flyby, orbiter, lander, probe & Undifferentiated, metamorphosed asteroids & Low \\
\hline
II & All outbound types: flyby, orbiter, lander, probe & Venus, Moon, comets, carbonaceous chondrite asteroids, Titan, Jupiter, Saturn, Pluto, Kuiper Belt Objects & Low \\
\hline
III & No direct contact: flyby, orbiter & Mars, Europa, Enceladus & Low \\
\hline
IV & Direct contact: lander, probe & Mars, Europa, Enceladus & High \\
\hline
V(r) – restricted & Earth return & Mars, Europa, Enceladus & High \\
\hline
V(u) – unrestricted & Earth return & Venus, Moon, asteroids, comets & N/A \\
\hline
\end{tabular}
\end{center}

\textbf{Table 1:} COSPAR's mission categorization {[}10,12{]}

\vspace{1\baselineskip}
Category I is the least problematic one: undifferentiated, metamorphosed asteroids are primordial remnants from the early formation of the solar system, and for this reason, they do not bear significance concerning the search for life in the solar system. Category II, on the contrary, presents the most difficult case. This category encompasses

\vspace{1\baselineskip}
\noindent "[...] all types of missions to those target bodies where there is significant interest relative to the process of chemical evolution and/or the origin of life, but where scientific opinion provides a remote chance of contamination by organic or biological materials which could compromise future investigations of the process of chemical evolution and/or the origin of life." [12]

\vspace{1\baselineskip}
\noindent One can find under the umbrella Category II a very broad range of missions: deep space probes such as Voyager I and II or New Horizons, orbiters of gas and icy giants, Venus or Mercury landers, as well as return sample missions from carbonaceous chondrite asteroids, comets, and most of the planetary satellites in our solar system, the Earth’s Moon included. The common denominator of all these missions is that although they may include close approach to or direct contact with extraterrestrial environments, there is only a very low chance that any members of the terrestrial biosphere may survive or even replicate there, in turn posing threats to future chances of discovering organic material of extraterrestrial origin on these sites [12]. However, given that the consensus on the survivability of terrestrial organisms in extreme space conditions keeps evolving, the precise boundaries of Category II are very hard to define [13]. For example, Titan is listed among the target bodies in this category, but given that astrobiologists hypothesize an alternative biosphere model on Titan using liquid methane instead of H2O as the principal driver of chemical recombination, it is hard to assess how grave the risk of contamination by terrestrial life is here [14]. Down the line, this boils down to the problem with defining life as such, which haunts any attempt to investigate the origin and chemical evolution of life on Earth as well as in space. 

With Categories III and IV, we enter the high-stakes territory of celestial bodies that may have boasted life in the past or may even contain surviving lifeforms in the present. In their case, “scientific opinion provides a significant chance of contamination by organic or biological materials” [12]. The high level of precaution here seems indeed warranted: even if no landing is intended—as in Category III missions—the measures should be implemented to rule out the chances that biological contamination happens unwittingly, e.g. in an accidental crash of an orbiter. Elimination of contamination is then implemented by a series of procedures including use of cleanrooms during the process of assembling the machinery and monitoring the residual inventory of organic materials in the spacecraft—most importantly the hardware’s \emph{bioburden} measured in so-called \emph{colony forming units} (CFU/m2) of bacterial spores that can survive spaceflight in a dormant state and remain inactive for millennia [10]. In line with these procedures, NASA’s Office of Planetary Protection (which evolved out of the original Planetary Quarantine Office that held the agenda during the Apollo missions) formulated a planetary protection standard (NASA-STD-8719.27) that specifies maximal permissible bioburden levels and precautionary steps to be taken for different kinds of subcategories within Category III and IV. For example, it requires that the number of spores at launch of a Category IV Mars-bound spacecraft should be “less than 5.0 x 105 total spores on all landed hardware,” [10] while in other cases, it formulates new numerical criteria: safe exploration of icy moons such as Europa or Enceladus demands “less than 1 × 10-4 probability that a biological inoculation event occurs to a habitable environment on these worlds” [10].

Finally, for the return sample missions with both outbound and inbound legs of the journey, the Category V classification first requires determining whether the given target body of the mission’s outbound phase is “deemed by scientific opinion to have no indigenous life forms” [12]. If that is the case, the mission falls within Category V(u)—unrestricted Earth return—which means that no additional specific measures should be implemented for the inbound leg of its journey. However, should there be no consensus on whether the target body harbours indigenous forms of life (e.g. Mars, Europa),

\vspace{1\baselineskip}
\noindent [...] "the highest degree of concern is expressed by the absolute prohibition of destructive impact upon return, the need for containment throughout the return phase of all returned hardware which directly contacted the target body or unsterilized material from the body, and the need for containment of any unsterilized sample collected and returned to Earth. Post-mission, there is a need to conduct timely analyses of any unsterilized sample collected and returned to Earth, under strict containment, and using the most sensitive techniques. If any sign of the existence of a non-terrestrial replicating entity is found, the returned sample should remain contained unless treated by an effective sterilizing procedure." [12]

\vspace{1\baselineskip}
\noindent This high level of concern implies that a given mission falls into Category V(r), although no past return sample mission has yet been categorized as such. Out of all the past missions, the target bodies were only those that fall within the Category II for the outbound leg, and hence classify as Category V missions with unrestricted Earth return: Moon (6 crewed Apollo missions, Soviet robotic missions Luna 16, Luna 20 and Luna 24, Chinese probes Chang’e 5 and 6), asteroids (JAXA’s Hayabusa and Hayabusa2, NASA’s OSIRIS-REx) and comets (Stardust). Moreover, Genesis mission can be counted among return sample missions, although its purpose was not to collect samples of another celestial body, but of the solar wind particles, and there is currently one asteroid sample return mission in progress: China’s Tianwen 2. The first chapters in the history of Category V(r) missions are thus going to be written only in the years to come, most prominently by the Mars Sample Return (MSR) mission: the joint campaign by NASA and ESA to bring back samples currently collected by the Perseverance rover in the Jezero crater on Mars. Simultaneously, China plans its own return sample mission to Mars (Tianwen 3), and Japan prepares for the launch of MMX (Martian Moons eXploration), scheduled to bring back samples from the largest moon of Mars—Phobos—by the end of 2031. 

Since all Category V missions circumscribe a full loop between terrestrial and extraterrestrial environments, they are extremely relevant to earth-space governance. Even if this loop—represented by the outbound journey followed by the in-site sample collection and the inbound flight back home—is not present, it is warranted to think about the governance of outer space exploration from the earth-space governance perspective. Planetary protection policies demonstrate this, as they operationalize human awareness of the potential harmful intervention into extraterrestrial environments, although they narrowly focus on the issue of biological contamination from the perspective of the instrumental value the extraterrestrial sites have for the scientific understanding of the origins, evolution, and diversity of lifeforms. Despite this limitation, planetary protection may be treated as the expanded application of ecology, to the extent to which it assesses potential interactions between terrestrial lifeforms with extraterrestrial environments and vice versa. Thus, it aligns well with the assumptions of earth-space governance by considering “space as an integral part of ecological and social concerns,” [2] which is also emphasized by Galli and Losch’s discussion of “planetary sustainability” in an explicitly cosmic viewpoint [8]. Ecology is always already a multi-planetary affair: the future of life on Earth, including issues of terrestrial habitability in general and the survivability of the Earth’s environment by humans in particular is tied to how we understand the past, present, and future of life in the universe, and it follows that the governance and policy frameworks should adopt this cosmic standpoint as well. Integrating outer space exploration into the larger human endeavour of safeguarding sustainability on the earth-space axis thus seems to be one of the logical next steps in developing earth-space governance approaches. The question to be answered then is how this integration should proceed.

\section{Conservationist paradigm}

One of the ways to integrate outer space exploration into the
earth-space governance is to follow the planetary protection's emphasis
on the issues of biological contamination, thus focusing on the
questions of \emph{earth-space sustainability} {[}15{]}. This may lead
to the conclusion that planetary protection encourages the treatment of
extraterrestrial sites as protected areas that should remain maximally
intact, especially to avoid the overspill of terrestrial microbes that
would severely compromise the search for life outside Earth. In analogy
with terrestrial environments, one can thus envision wilderness
protection policies for planetary bodies, as proposed by Charles Cockell
and Gerda Horneck {[}6{]}. According to these authors, the idea of
\emph{planetary parks} operationalizes the UN Outer Space Treaty's and
COSPAR's focus on preventing biological contamination for the sake of
future space exploration. Additionally, their arguments open more
avenues for the justification of the concept of planetary park, such as
an argument from the intrinsic value of extraterrestrial environments
(i.e. portraying their value as irreducible to human instrumental
goals), as well as alternative utilitarian reasons transcending issues
of scientific exploration (e.g. aesthetic value or pedagogical value to
future humans).

Cockell and Horneck ground their approach in environmental ethics, which
emphasises \emph{land} as a kernel of various instrumental or intrinsic
values. A similar perspective is taken up by Woodruff T. Sullivan III
{[}7{]}, who models his \emph{planetocentric ethics} after the
environmentalist classics such as Aldo Leopold's \emph{land ethics} or
Arne Naess' \emph{deep ecology}, who all argue for the moral standing of
non-human entities irreducible to how they can be instrumental to
humans. By extension, Sullivan argues, celestial bodies have intrinsic
value, since they represent unique entries in the diverse inventory of
the solar system and should be accordingly treated with maximum care and
protection. The maxim of planetocentric ethics then reads as follows:
``Cause neither physical nor biological harm to any planetary body and
its ecosystems'' {[}7{]}. Sullivan's justification of this maxim relies
on the assumption he borrows from Holmes Rolston III that every
celestial body is an irreplicable ``nature's project'', and any
artificial intervention can only spoil the unfolding of such a project
{[}16{]}. As Sullivan argues, this situation is analogous to the value
of biological species in the terrestrial environment, where ``every
species has intrinsic value as a part of the whole,'' {[}7{]} and
therefore requires special protection. From this standpoint, any
nature's project supersedes human imperfect judgment about the project's
apparent purposes, and hence it must be shielded from human interference
that is portrayed as necessarily harmful. This assumption is mobilized
in Sullivan's refutation of terraforming Martian environments:

\vspace{1\baselineskip}
\noindent "My central concern is that we humans simply do not have (nor are likely to acquire at any time in the foreseeable future) the requisite knowledge and wisdom to “help” an ecosystem or planet gain its “fullest potential”— we are not even good at this kind of engineering on Earth, let alone with an exotic ecosystem. If we were to carry out such modifications, we would irreparably contaminate “nature’s project” on Mars, so that it would be lost to scientific study forever. We gain more by observing nature’s planetary experiments than by conducting our own, especially those that are irreversible." [7]

\vspace{1\baselineskip}
The echoes of this approach can also be found in other intrinsic value
justifications of biological contamination prevention, such as
\emph{biocentric ethics}. Applied to the case of extraterrestrial
environments by Michael Mautner {[}17{]}, life has, in his account, a
unique position in the cosmos, thanks to its simultaneous complexity and
fragility. The natural purpose of life is its perpetuation---the
propagation of genetic material across generations---and for that
reason, it must be preserved against many endogenous and exogenous
threats that may lead to breakdown of its complex nature. In analogy to
the purported value of life, geological features are sometimes proposed
to require the same kind of treatment, too, hence gesturing towards the
protection of the cosmic diversity of \emph{exogeosites} (i.e.
geologically important sites outside the Earth). Explicitly named as
\emph{exogeoconservation}, this approach mandates ``the identification
of scientific, historic, aesthetic, ecological or cultural value in
celestial bodies and their component geological and geomorphological
features, and the protection of such bodies and features'' {[}18{]}.

Interestingly, all these reviewed proposals refer to terrestrial precedents to establish the plausibility of their approach, such as the US Wilderness Act or the Antarctic Treaty [6,7,18]. These examples also serve to the authors as a proof that the implementation of their ethical standards is feasible, albeit quite demanding. As Sullivan admits, “planetocentric ethics will not be an easy ethical scheme to put into force,” but because “the Solar System is our last great wilderness,” [7] the imperative is to act pre-emptively and maximize the protection of the extraterrestrial sites while the human intervention is still relatively minuscule, compared to what technologically more capable future generations may be able to do. Hence, through this emphasis on the notion of wilderness, what connects all these proposals is the gesture of environmental conservation: the extraterrestrial sites shall be left maximally intact, ideally shielded from any influence of terrestrial origin, whether directly or indirectly anthropogenic (e.g. crewed mission vs. robotic exploration). This conservationist paradigm thus proceeds with identifying celestial bodies as primary units of care, towards which humans as moral agents have ethical obligations. By virtue of doing so, the conservationist paradigm also identifies a special connection between life and celestial bodies that harbour it: a planet or a moon is the primary unit to which life belongs. This perspective seems to be consistent with the prevalent views on the origins of life, where planets play the crucial role of spinning incubators of lifeforms [19].

\section{Constructivist paradigm}

The trouble with the conservationist paradigm begins in the moment when the boundaries between entities deemed to be separated from each other start to collapse; the conservationist approach works well only if we treat celestial bodies as separate environments—a kind of curved, isolated megalandscapes. However, such a simple analogy with land—understood as a clearly delineated territory to be protected—is impossible to sustain; planets are not bloated versions of land maximized to a nearly round shape and isolated from whatever lies beneath by some clear demarcation line. They are hard to isolate from their environment in terms of both their history and dynamics. The analogy with the protection of biodiversity also falls short: planets are singular cosmic bodies, not a self-reproducing, genetically related population/species.

The first thing to note in this respect is that extraterrestrial environments spontaneously cross-contaminate, and this observation can be used to construct an intuitive litmus test for the soundness of the conservationist paradigm. One famous example of a natural transport of material between celestial bodies represents the case of Allan Hills meteorite (ALH84001), originally found in Antarctica in 1984, which turned out to be a piece of Martian rock carrying traces of what was believed to be nanofossils of indigenous Martian microorganisms [20,21]. Although the evidence for such extraordinary claims proved to be far from conclusive, ALH84001 nicely illustrates that celestial bodies can potentially contaminate each other also without human influence, which problematizes the idea that contamination of extraterrestrial environments is an absolute no-go for space exploration, too: one may argue that if biological contamination is in some cases a “nature’s project”, moral agents may simply follow nature’s lead here. 

Woodruff T. Sullivan III recognises this trouble in his discussion of
planetocentric ethics. To counter the case of natural
cross-contamination, he claims that these kinds of environmental
influences that spontaneously play out between celestial bodies do not
constitute a violation of the planet's intrinsic value, while human
influence inevitably does so: he portrays human influence not as a
continuation of ``nature's projects'' by other means, but as an
\emph{interference} with the unfolding of nature's plans. Such
conservationist argumentation is, however, highly problematic because it
separates human activity from the rest of the processes happening in the
solar system. This approach is not metaphysically warranted, since---as
emphasised by many planetary philosophers---celestial bodies are highly
porous, dynamic entities that constantly renegotiate their separation
from their cosmic surroundings {[}22,23{]}. The transport of materials
via meteorites between Mars and Earth is but one example of this
porosity: the meteoric/comet bombardment of planets in early solar
system (believed to be responsible for importing the organic molecules
necessary for later bootstrapping of terrestrial biochemistry), the
collisions between planetary bodies (such as the crash of a proto-planet
Theia into young Earth that resulted in formation of Moon), the
electromagnetic gusts of solar wind, or the gravitational influence
between planets make the case even stronger.

One may even say that the core of the astronomical definition of the
planet lies in highlighting this dynamic, processual nature of planets,
implying how environmental factors such as the relative gravitational
dominance of a planet influence whether it really can be categorized as
a planetary body {[}24,25{]}. Planets are evolving accretions of cosmic
matter that formed in the volatile environment of their nascent solar
systems, and this \emph{expanded ecology} remains the principal
condition of their integrity and existence over their entire history
{[}5, 26{]}. By extension, it is the solar system, rather than a planet,
which may be treated as the primary unit that life belongs to,
since---in the last instance---the solar system is the primary site of
causal forces necessary for life's emergence and maintenance, starting
with the sovereign influence of the host star as the main source of
energetic inputs for the planetary energy gradients necessary for any
thinkable lifeform {[}27{]}. Planetary life is thus impossible without
the extraplanetary forces that shape the environment hospitable to
life's origins.

Seeing planets as involutions of cosmic exteriority rather than as
territories separated from the antagonistic, contaminating outside
shatters the conservationist approach, since the terms of what is to be
conserved and what the contaminating agents are must be constructed and
justified anew in each case. To support this shift in perspective, one
may think about the implications of the relative abundance of basic
organic compounds such as amino acids on comets and asteroids---the
knowledge we possess mainly thanks to the return sample missions. The
near-Earth asteroid Bennu visited by OSIRIS-REx mission brought back
samples of the asteroid's regolith that contain 14 out of 20 amino acids
used in terrestrial organic chemistry, as well as the 5 nucleobases of
DNA and RNA (adenine, cytosine, guanin, thymine and uracil) {[}28{]};
similar results concerning the presence of amino acids as well as
phosphorus have been obtained from the samples of asteroid Ryugu
delivered by JAXA's Hayabusa2 mission {[}29{]}. Comets have been shown
to have similarly complex composition when it comes to the elements of
organic chemistry: for example, the analysis of
67P/Churyumov-Gerasimenko---the comet studied by ESA's Rosetta
mission---revealed that its coma (the comet's temporary atmosphere that
forms as the object approaches the Sun, creating the characteristic
``tail'' of the comet) features three simple amino acids (glycine,
methylamine, ethylamine) {[}30{]}.

All these findings have been subject to an ongoing debate about the
potential extra-terrestrial sources of organic chemistry on Earth,
suggesting that the direct genealogy of life as we know it here on Earth
goes deep into pre-planetary and proto-planetary history {[}31{]}. Early
Earth was bombarded by comets and meteorites that could have delivered
the basic building blocks of life, and in concert with the energetic
gradients present on the planet (e.g. around the active hydrothermal
vents in young oceans), generated a well-stirred-and-shaken substance
from which terrestrial biology emerged {[}32{]}. Some of the amino acids
most probably formed in the very earliest stages of the solar system
formation, when the proto-planetary objects known as planetesimals
coalesced in the disc around the Sun, and by bumping into each other,
accreted into larger structures that eventually became the planets. Many
asteroids and comets are the leftover material from these early stages
of solar system formation, and thus they stand for well-preserved
fossils that let us peek into the composition of the ``construction
material'' available in the early solar system.

At this point, the saying that we are all made of stardust ceases to be
a metaphor. Rather than pristine wilderness, the perspective of the
shared, cosmic origin of life invites instead an image of the solar
system as a \emph{construction site}: a project of evolving cosmic
matter, with planets and biogeochemistries as its products. The major
difference from the conservationist ``nature's project'' approach lies
here in treating human exploration of the solar system as the
constructive continuation of the big history of the evolution of life in
their home solar system;~humans and their actions elongate cosmic
projects that have abiotic, impersonal, elementary, even pre-planetary
origins. The assumption of a superior plan hidden in natural processes,
a plan that automatically supersedes human judgement, seems to rule out
any possibility that at the more-than-planetary scale, human species can
do anything beyond being a passive observer of the virtuous unfolding of
``nature's projects''---an assumption that seems to contradict the vital
dynamisms inherent in the constructive process of the solar system's
formation, evolution, and maturation.

The final argument to be made in favour of this ``constructivist''
paradigm (as opposed to the conservationist, see \textbf{Table 2}) is
related to the uncertainty about what exactly counts as a living
structure. In recent years, astrobiologists started questioning
definitions of life that rely too much on specific conditions of
terrestrial biology, such as the presence of liquid water on the surface
or the key chemical elements: carbon, hydrogen, nitrogen, oxygen,
phosphorus, and sulphur (CHNOPS) {[}33{]}. Motivated by uncanny
environments of Jupiter's and Saturn's moons, such as Titan, Enceladus,
or Europa, multiple alternative biologies have been proposed, including
biochemistries based on silicon instead of carbon as the key
compositional element, or those thriving in frozen seas of ammonia or
ethane on cold, Titan-like worlds {[}34{]}. Beyond the confines of the
search for life in the solar system, even weirder forms of life were
proposed, e.g. mechanothrops using motor-like modules to harvest kinetic
energy {[}35{]}, or advanced intelligent lifeforms that fully abandoned
organic substrates in favour of presumably more durable machinic,
technological forms of existence (the \emph{postbiological evolution}
hypothesis {[}36{]}). Moreover, if all life in the solar system shares a
pre-planetary origin in the organic material assembled in the accretion
disk and carried by comets or asteroids until today, this common genesis
problematizes the strict distinction between isolated lineages of
lifeforms that should be kept separated at all costs. Hence, as the
definition of life becomes looser and more inclusive, the concept of
biological contamination may become impossible to sustain, because it
will be very difficult to distinguish what exactly we are protecting
from what.

\begin{center}
\renewcommand{\arraystretch}{1.5}
\noindent 
\begin{tabular}{|p{3cm}|p{5cm}|p{5cm}|}
\hline
\textbf{} & CONSERVATIONIST PARADIGM & CONSTRUCTIVIST PARADIGM \\
\hline
\textbf{Core metaphor} & Wilderness & Construction site \\
\hline
\textbf{Unit of care} & Planet (or other celestial body) & Solar system (expanded ecology) \\
\hline
\textbf{Ontology} & Discrete, static objects & Porous, dynamic processes \\
\hline
\textbf{View of humanity} & Contaminant, intruder & Directly involved agent \\
\hline
\textbf{Ethical imperative} & Non-interference & Continuation of construction \\
\hline
\textbf{View of nature/cosmos} & Separate from humanity & A continuum that includes humanity \\
\hline
\end{tabular}
\end{center}

\textbf{Table 2:} Schematic comparison of conservationist and constructivist paradigms

\section{Discussion: Conservation as a mode of construction}

How does this tentative constructivist paradigm vibe with the existing
exploration protocols and future mission plans? Before answering this
question, let me first clarify that it would be a mistake to assume that
the motivation of the argumentation presented in this paper is to refuse
conservation \emph{tout court}; instead, the point is to emphasise that
the conservation itself is a mode of action that can be subsumed under
the umbrella term of constructive processes. Hence, instead of pitting
conservation against construction as direct opposites, the
constructivist paradigm is rather geared to pragmatically appreciate a
whole portfolio of forms of actions, among which conservation figures as
but one kind among many others. In other words, there may still be good
reasons for pursuing conservationist agendas in many cases, despite the
justification for these agendas stems from a different set of
assumptions than those shared among the authors in the conservationist
paradigm. To give you a simple example, consider a case of nature
reserves on Earth: they are artificial constructions, deliberate design
plans, and their pristine nature is the function of the human
intervention that conserves them in a certain state via legislation and
enforcement apparatus. Similarly, on a much grander scale, E. O.
Wilson's Half-Earth is a project of deliberate construction of
artificial wilderness through the strategy of human withdrawal {[}37{]}.
Rather than being passively observed, they are actively maintained, and
the same may hold for different regions of the solar system, identified
as valuable for one pragmatic reason or another. Conservation \emph{is}
construction.

If there is one line of argumentation where the proponents of the
conservationist paradigm may agree with the constructivist viewpoint, it
is the issue of the feasibility of restrictive protocols for engaging
with extraterrestrial environments. Christopher P. McKay, a planetary
scientist at NASA's Ames Research Center and one of the prominent
Martian astrobiologists, claims that ``the cost for sterile exploration
is high, and many argue that the low probability of life does not
justify this cost. In addition, sterile exploration precludes human
exploration, and it is likely that human exploration is needed to
adequately resolve the question of life'' {[}38{]}. Sooner or later, in
his opinion, the biological contamination protocols will break down
under the weight of the increasing interest in space exploration, and
for that reason, he suggests the following pragmatic approach:

\vspace{1\baselineskip}
\noindent The question of the future of life in the Universe involves science but also involves human choice at a basic level. Thus it is important for us to consider what purposes and values will guide our choices. I suggest that the long-term goal for astrobiology be to enhance the richness and diversity of life in the Universe. Clearly this begins with maintaining the richness and diversity of life on Earth. Next comes the search for life beyond Earth to give us an understanding of the natural diversity of life. Third, we can investigate the potential for life from Earth to spread beyond. {[}38{]}

\vspace{1\baselineskip}
\noindent With respect to the main proclaimed goal---``to enhance the richness and diversity of life in the Universe''---the argumentation in the previous section prepared the ground for the statement that construction may be a default mode of action life does (through propagating itself), and humans can be carriers of this constructive endeavour further by modifying extra-terrestrial environments for the sake of diversification of the life's lineages. Even if Mars crewed missions, permanent Mars settlements, or the terraforming of Mars may be very distinct and perhaps unattainable objectives, the very presence of technical beings of terrestrial origin on other celestial bodies in the solar system is a testament to life's capacity to propagate its manifestations beyond its home planet. From Mars rovers and landers through asteroid probes to future missions on the surface of Jovian or Saturnian moons (e.g. the poetically named Dragonfly, the first rotorcraft to study Titan) {[}39{]}, these are the post-biological emissaries of the Earth's biosphere, the offspring of the solar system's life-making capacities. Perhaps, by means of robotic exploration alone, humans are already positively enhancing the habitability of Mars or other celestial bodies, since their environments now feature very complex entities that would not otherwise be there---the machinic lifeforms {[}40{]}.

Concerning return sample missions such as Mars Sample Return (MSR), the
constructivist approach can be reinforced by treating the technical
artefacts as emissaries of the terrestrial biosphere, not as dangerous
intruders potentially contaminating other celestial bodies. Yes, the
biological contamination is major issue, since there is a huge pragmatic
value in pre-emptively acting in favour of preserving hypothetical
indigenous lifeforms, so that they can be potentially discovered at the
first place, but this imperative against contamination is driven by
purely scientific rationale---tampering with a sensitive environment
inevitably destroys the scientific gains of solar system exploration.
Yet, down the line, the value of interventions that can enhance cosmic
ecosystems in terms of their biodiversity can trump the continuous
importance of conservation. For example, McKay proposes to attempt
growing plants on Mars using Martian soil in biologically isolated
containers {[}41{]}, and different cases for artificially induced,
non-human ``wilding'' of Mars have also been proposed by the artist
Alexandra Daisy Ginsberg {[}42{]}. When it comes to future asteroid
missions, Charles Cockell's projects, such as BioAsteroid and BioRock,
pave the way to using extraterrestrial environments as life
laboratories, enabling scientists to study the interaction between
cosmic material and organic structures in outer space conditions
{[}43{]}. His proposals can be embraced, especially in the case of
future Category I and II missions, where the biohazards are relatively
low and remote: thanks to their abundance, richness, and relative
isolation, asteroids can become sites of astrobiological
experimentation.

In the future decades, the planetary protection approach thus should be
guided by pragmatic protocols supporting specific scientific goals of
missions and institutionalized research programs, rather than by
absolute moral imperatives: it should be more about \emph{handshake
protocols} guiding the responsible on-ramp of deeper engagement between
celestial bodies, and less about containment or isolation, since the
ultimate value of the outer space exploration lies in the constructive
continuation of the solar system's evolution. Here, earth-space
governance meets science governance: the question of safeguarding
earth-space sustainability becomes a question of responsible,
sustainable scientific exploration that is fully aware of the
constructive role of humans in solar system ecology. Solar system
exploration is an exploration of one big ecosystem sharing one genealogy
of origins, both in the case of the celestial bodies and in the case of
the organic bodies that inhabit them.

\section{Conclusion}

This paper has charted a course from the conservationist planetary
protection paradigm to a new constructivist approach applicable to the
issues of earth-space governance, while remaining closely aligned with
NASA's or COSPAR's policies. The core of this paradigm lies in a
perspectival rotation from viewing planets as discrete, isolated
wilderness to understanding them as dynamic, porous involutions of the
expanded solar system ecology. The conservationist paradigm, with its
terrestrial metaphor of pristine nature and its ethical imperative of
non-interference, fails to recognize the dynamic nature of astronomical
bodies. The constructivist paradigm, in contrast, embraces it. By
recognizing that Earth\textquotesingle s biosphere is itself a product
of cosmic ingredients and that humanity is a constructive agent within
this ongoing process of the solar system's evolution, it offers a more
philosophically sound and pragmatically viable foundation for
earth-space governance that adds a new dimension of polemics to the
debate about human expansion to the cosmos. In particular, instead of
viewing the outer space as a new domain that the issues of Earth-system
governance should be expanded to, it shows that the cosmic reality is
the primary site according to which our governance paradigms should be
modelled; earth-space governance can be better named as
\emph{space-to-earth governance}: Earth is always already a cosmic body
\emph{in} outer space. The essential take-away of the constructivist
framework then may be much more about learning to implement the outer
space approach to Earth (and the issues of sustainability here and now
on the ground), rather than replicating Earth-centric intuitions in the
issues of space governance.

In terms of practical implications, the constructivist paradigm does not
call for the abandonment of caution. Rather, it ramifies planetary
protection as a pragmatic, evolving system of risk management. The
problem can no longer be treated in absolute terms: The question
``Should we risk contamination?'' shall be better formulated as ``What
are the specific, scientifically evaluated risks and benefits of a
particular interaction on a particular world?'' This means that the
issue of biological contamination for Category V(r) missions like the
Mars Sample Return (MSR) is still treated with utmost care, but the
underlying rationale shifts. The extreme containment protocols are not
justified by a fear of violating a pristine world, but as a rational
management of an unknown, albeit low-probability, risk to Mars's
biosphere. The priority is given to risk management over \emph{risk
aversion}, guided by theory-informed, evidence-based prudence over
absolute prohibition {[}44, 45{]}.

The constructivist view also invites a re-evaluation of humans as
constructive agents in their home solar system. Humans---alongside the
terrestrial biosphere they originate from---are not ``contaminant''
species to be contained, but an expression of the solar system's
creative capacities. However, this does not mean to side with the
escapist fantasies of abandoning the ruined Spaceship Earth for a
pristine landscapes of Mars (as the second chance for human species to
``get things right''); such a view would ignore the fact that
environmental problems on Earth are the results of human failure to
manage the habitability of their planetary home---a decision-making
failure humans shall inevitably carry anywhere else they go as a looming
threat. Planetary stewardship on our home turf is not a separate issue
from space exploration; it is the foundational expression of the
responsibility towards the continuation of life in our solar system. The
constructivist paradigm acknowledges that, as moral agents, humans
cannot escape the responsibility of making value judgments and measuring
the consequences of their actions. This position is more philosophically
honest and ethically robust, recognizing that our cognitive abilities
grant us not a right to colonize, but a duty to choose: a duty to carry
on the difficult, necessary, and creative work of responsible
construction that started 4.5 billion years ago.

\section*{Acknowledgements}
The work on this paper has been supported by the project funded by the Scientific Grant Agency of the Ministry of Education of Slovak Republic - VEGA no. 2/0110/24 “Tasks of Political Philosophy in the Context of the Anthropocene II”

\section*{Author bio}
During the preparation of this work the author used Gemini 2.5 Pro to improve the readability of the Section 1 and Section 6 of the article. After using this tool, the author reviewed and edited the content as needed and takes full responsibility for the content of the publication.

\section*{References}
\textcolor{white}{jjjji}{[}1{]} T. Harrison, T. Nahmyo. NASA in the Second Space Age:
Exploration, Partnering, and Security. \emph{Strategic Studies
Quarterly} 10, 4 (2016), 2--13.

{[}2{]} X.-S. Yap, R. E. Kim. Towards Earth-Space Governance in a
Multi-Planetary Era. \emph{Earth System Governance} 16 (April 2023),
100173. \url{https://doi.org/10.1016/j.esg.2023.100173}.

{[}3{]} D. Chakrabarty. \emph{The Climate of History in a Planetary
Age}. The University of Chicago Press, 2021.

{[}4{]} N. Clark, B. Szerszynski. \emph{Planetary Social Thought: The
Anthropocene Challenge to the Social Sciences}. Polity Press, 2021.

{[}5{]} L. Likav\v{c}an. Another Earth: An astronomical concept of the planet for the environmental humanities. \emph{Distinktion: Journal of Social Theory} 25, 1 (2024), 17--36. \url{https://doi.org/10.1080/1600910X.2024.2326448}.

{[}6{]} C. S. Cockell, G. Horneck. Planetary Parks---Formulating a
Wilderness Policy for Planetary Bodies. \emph{Space Policy} 22, 4
(2006), 256--61. \url{https://doi.org/10.1016/j.spacepol.2006.08.006}.

{[}7{]} W. T. Sullivan III. Planetocentric Ethics: Principles for
Exploring a Solar System That May Contain Extraterrestrial Microbial
Life. In \emph{Encountering Life in the Universe: Ethical Foundations
and Social Implications of Astrobiology}, edited by W. R. Stoeger, A.H.
Spitz, C. Impey. University of Arizona Press, 2013.

{[}8{]} A. Galli, A. Losch. Beyond Planetary Protection: What Is
Planetary Sustainability and What Are Its Implications for Space
Research? \emph{Life Sciences in Space Research} 23 (November 2019),
3--9. \url{https://doi.org/10.1016/j.lssr.2019.02.005}.

{[}9{]} D. J. Des Marais, J. A. Nuth, L. J. Allamandola, et al. The NASA
Astrobiology Roadmap. \emph{Astrobiology} 8, 4 (2008), 715--30.
\url{https://doi.org/10.1089/ast.2008.0819}.

{[}10{]} NASA. \emph{NASA Planetary Protection Handbook}.
NASA/SP-20240016475, Version 1.0. National Aeronautics and Space
Administration, 2024.

{[}11{]} UN General Assembly. Outer Space Treaty. 1967.
\url{https://www.unoosa.org/oosa/en/ourwork/spacelaw/treaties/outerspacetreaty.html}.

{[}12{]} Committee on Space Research. COSPAR Policy on Planetary
Protection. \emph{Space Research Today}, 220 (July 2024), 14--36.

{[}13{]} N. Merino, H. S. Aronson, D. P. Bojanova, et al. Living at the
Extremes: Extremophiles and the Limits of Life in a Planetary Context.
\emph{Frontiers in Microbiology} 10 (April 2019), 780.
\url{https://doi.org/10.3389/fmicb.2019.00780}.

{[}14{]} L. H. Norman. Is There Life on \ldots{} Titan? \emph{Astronomy
\& Geophysics} 52, 1 (2011), 1.39-1.42.
\url{https://doi.org/10.1111/j.1468-4004.2011.52139.x}.

{[}15{]} X.-S. Yap, B. Truffer. Contouring `Earth-Space Sustainability.'
\emph{Environmental Innovation and Societal Transitions} 44 (September
2022), 185--93. \url{https://doi.org/10.1016/j.eist.2022.06.004}.

{[}16{]} H. Rolston III. The Preservation of Natural Value in the Solar
System. In \emph{Beyond Spaceship Earth: Environmental Ethics and the
Solar System.}, edited by E. C. Hargrove. Sierra Club Books, 1986.

{[}17{]} M. N. Mautner. Life-Centered Ethics, and the Human Future in
Space. \emph{Bioethics} 23, 8 (2009), 433--40.
\url{https://doi.org/10.1111/j.1467-8519.2008.00688.x}.

{[}18{]} J. J. Matthews, S. McMahon. Exogeoconservation: Protecting
Geological Heritage on Celestial Bodies. \emph{Acta Astronautica} 149
(August 2018), 55--60.
\url{https://doi.org/10.1016/j.actaastro.2018.05.034}.

{[}19{]} Spitzer, Jan. \emph{How Molecular Forces and Rotating Planets
Create Life\,: The Emergence and Evolution of Prokaryotic Cells}. The
MIT Press, 2021.

{[}20{]} S. J. Dick. \emph{Space, Time, and Aliens: Collected Works on
Cosmos and Culture}. Springer, 2020.

{[}21{]} D. S. McKay, E. K. Gibson, K. L. Thomas-Keprta, et al. Search
for Past Life on Mars: Possible Relic Biogenic Activity in Martian
Meteorite ALH84001. \emph{Science} 273, 5277 (1996), 924--30.
\url{https://doi.org/10.1126/science.273.5277.924}.

{[}22{]} B. Szerszynski. Planetary Mobilities: Movement, Memory and
Emergence in the Body of the Earth. \emph{Mobilities} 11, 4 (2016),
614--28. \url{https://doi.org/10.1080/17450101.2016.1211828}.

{[}23{]} N. Clark, B. Szerszynski. What Can a Planet Do? \emph{Cultural
Geographies} 32, 3 (2025), 331--41.
\url{https://doi.org/10.1177/14744740251326904}.

{[}24{]} M. L. Wong, M. Duckett, E. S. Hernandez, V. Rajaei, K. J.
Smith. The Process We Call Earth: Relationships Between Dynamic
Feedbacks and the Search for Gaiasignatures in a New Paradigm of
Earthlikeness. \emph{Perspectives of Earth and Space Scientists} 5, 1
(2024), e2023CN000223. \url{https://doi.org/10.1029/2023CN000223}.

{[}25{]} S. Soter. What Is a Planet? \emph{The Astronomical Journal}
132, 6 (2006), 2513--19. \url{https://doi.org/10.1086/508861}.

{[}26{]} G. Basri, M. E. Brown. Planetesimals To Brown Dwarfs: What Is a
Planet? \emph{Annual Review of Earth and Planetary Sciences} 34, 1
(2006), 193--216.
\url{https://doi.org/10.1146/annurev.earth.34.031405.125058}.

{[}27{]} A. Kleidon. Life, Hierarchy, and the Thermodynamic Machinery of
Planet Earth. \emph{Physics of Life Reviews} 7, 4 (2010), 424--60.
\url{https://doi.org/10.1016/j.plrev.2010.10.002}.

{[}28{]} D. P. Glavin, J. P. Dworkin, C. M. O'D. Alexander, et al.
``Abundant Ammonia and Nitrogen-Rich Soluble Organic Matter in Samples
from Asteroid (101955) Bennu.'' \emph{Nature Astronomy} 9, 2 (2025),
199--210. \url{https://doi.org/10.1038/s41550-024-02472-9}.

{[}29{]} C. Potiszil, T. Ota, M. Yamanaka, et al. Insights into the
Formation and Evolution of Extraterrestrial Amino Acids from the
Asteroid Ryugu. \emph{Nature Communications} 14, 1 (2023), 1482.
\url{https://doi.org/10.1038/s41467-023-37107-6}.

{[}30{]} K. Altwegg, H. Balsiger, A. Bar-Nun, et al. Prebiotic
Chemicals---Amino Acid and Phosphorus---in the Coma of Comet
67P/Churyumov-Gerasimenko. \emph{Science Advances} 2, 5 (2016),
e1600285. \url{https://doi.org/10.1126/sciadv.1600285}.

{[}31{]} D. Glavin. Amino Acids in Asteroids and Comets: Implications
for the Origin of Life on Earth and Possibly Elsewhere. April 17, 2012.
\url{https://ntrs.nasa.gov/citations/20120009029}.

{[}32{]} A. C. Waajen, C. Lima, R. Goodacre, C. S. Cockell. Life on
Earth Can Grow on Extraterrestrial Organic Carbon. \emph{Scientific
Reports} 14, 1 (2024), 3691.
\url{https://doi.org/10.1038/s41598-024-54195-6}.

{[}33{]} M. L. Wong, S. Bartlett, S. Chen, L. Tierney. Searching for
Life, Mindful of Lyfe's Possibilities. \emph{Life} 12, 6 (2022), 783.
\url{https://doi.org/10.3390/life12060783}.

{[}34{]} M. Lingam, A. Leyb. \emph{Life in the Cosmos: From
Biosignatures to Technosignatures}. Harvard University Press, 2021.

{[}35{]} S. Bartlett, M. L. Wong. Defining Lyfe in the Universe: From
Three Privileged Functions to Four Pillars. \emph{Life} 10, 4 (2020), 4.
\url{https://doi.org/10.3390/life10040042}.

{[}36{]} M. M. Ćirković, R. J. Bradbury. Galactic Gradients,
Postbiological Evolution and the Apparent Failure of SETI. \emph{New
Astronomy} 11, 8 (2006), 628--39.
\url{https://doi.org/10.1016/j.newast.2006.04.003}.

{[}37{]} E. O. Wilson. \emph{Half-Earth: Our Planet's Fight for Life}.
Liveright, 2016.

{[}38{]} C. P. McKay. Astrobiology and Society. The Long View. In
\emph{Encountering Life in the Universe: Ethical Foundations and Social
Implications of Astrobiology}, edited by W. R. Stoeger, A. H. Spitz, and
C. Impey. The University of Arizona Press, 2013.

{[}39{]} J. W. Barnes, E. P. Turtle, M. G. Trainer, et al. Science Goals
and Objectives for the Dragonfly Titan Rotorcraft Relocatable Lander.
\emph{The Planetary Science Journal} 2, 4 (2021), 130.
\url{https://doi.org/10.3847/PSJ/abfdcf}.

{[}40{]} Walker, Sara Imari. \emph{Life as No One Knows It: The Physics
of Life's Emergence}. Riverhead Books, 2024.

{[}41{]} C. P. McKay. The Case for Plants on Mars. \emph{Acta
Horticulturae}, 642 (October 2004), 187--92.
\url{https://doi.org/10.17660/ActaHortic.2004.642.20}.

{[}42{]} A. D. Ginsberg. The Wilding of Mars. 2019.
\url{https://www.daisyginsberg.com/work/the-wilding-of-mars}.

{[}43{]} University of Edinburgh, School of Physics and Astronomy.
Launch of BioAsteroid Experiment to Space Station. School of Physics and
Astronomy, July 22, 2020.
\url{https://www.ph.ed.ac.uk/news/2020/launch-of-bioasteroid-experiment-to-space-station-20-07-22}.

{[}44{]} M. S. Race. Mars Sample Return and Planetary Protection in a
Public Context. \emph{Advances in Space Research} 22, 3 (1998): 391--99.
https://doi.org/10.1016/S0273-1177(98)00036-2.

{[}45{]} E. Romero, D. Francisco. The NASA Human System Risk Mitigation
Process for Space Exploration. \emph{Acta Astronautica} 175 (October
2020), 606--15. \url{https://doi.org/10.1016/j.actaastro.2020.04.046}.

\end{document}